\documentclass[conference,a4paper]{IEEEtran}
\usepackage{xcolor}
\usepackage{balance}
\usepackage{fix-cm} %

\ifCLASSINFOpdf
\usepackage[pdftex]{graphicx}
\else
\usepackage[dvips]{graphicx}
\fi

\usepackage{tikz}
\usepackage{pgfplots}
\usepackage{pgfplotstable}
\pgfplotsset{compat=1.18}
\usetikzlibrary{calc,decorations.markings,decorations.pathmorphing}

\usepackage{amsmath}
\usepackage{amsfonts}
\usepackage{amssymb}
\usepackage{xfrac}

\DeclareMathOperator*{\argmin}{arg\,min}

\ifCLASSOPTIONcompsoc
\usepackage[caption=false,font=normalsize,labelfont=sf,textfont=sf]{subfig}
\else
\usepackage[caption=false,font=footnotesize]{subfig}
\fi

\usepackage{url}
\usepackage{hyperref}

\usepackage{glossaries-extra}

\setabbreviationstyle[acronym]{long-short}

\glssetcategoryattribute{acronym}{nohyperfirst}{true}

\newacronym{rt}{RT}{Ray Tracing}
\newacronym{drt}{DRT}{Differentiable \gls{rt}}
\newacronym{tx}{TX}{transmitter}
\newacronym{rx}{RX}{receiver}
\newacronym{mpt}{MPT}{Min-Path-Tracing}
\newacronym{ad}{AD}{automatic differentiation}
\newacronym{socp}{SOCP}{second-order cone program}
\newacronym{bfgs}{BFGS}{Broyden-Fletcher-Goldfarb-Shanno}

\usepackage[commentmarkup=uwave]{changes}
\definechangesauthor[color=blue]{SL}

\usepackage{fancyhdr}
\begin{document}
\title{Fast, Differentiable, GPU-Accelerated\\Ray Tracing for Multiple Diffraction and Reflection Paths}

\author{\IEEEauthorblockN{
    Jérome Eertmans\IEEEauthorrefmark{1},
    Sophie Lequeu\IEEEauthorrefmark{1},
    Benoît Legat\IEEEauthorrefmark{1},
    Laurent Jacques\IEEEauthorrefmark{1},
    Claude Oestges\IEEEauthorrefmark{1}
  }                                     %
  \IEEEauthorblockA{\IEEEauthorrefmark{1}%
  ICTEAM, Université catholique de Louvain, Louvain-la-Neuve, Belgium, \href{mailto:jerome.eertmans@uclouvain.be}{jerome.eertmans@uclouvain.be}}
}

\maketitle

\thispagestyle{fancy}

\begin{abstract}
  We present a fast, differentiable, GPU-accelerated optimization method for ray path tracing in environments containing planar reflectors and straight diffraction edges. Based on Fermat's principle, our approach reformulates the path-finding problem as the minimization of total path length, enabling efficient parallel execution on modern GPU architectures. Unlike existing methods that require separate algorithms for reflections and diffractions, our unified formulation maintains consistent problem dimensions across all interaction sequences, making it particularly suitable for vectorized computation. Through implicit differentiation, we achieve efficient gradient computation without differentiating through solver iterations, significantly outperforming traditional automatic differentiation approaches. Numerical simulations demonstrate convergence rates comparable to specialized Newton methods while providing superior scalability for large-scale applications. The method integrates seamlessly with differentiable programming libraries such as JAX and DrJIT, enabling new possibilities in inverse design and optimization for wireless propagation modeling. The source code is openly available at \url{https://github.com/jeertmans/fpt-jax}.
\end{abstract}

\vskip0.5\baselineskip

\section{Introduction}

When determining possible ray paths connecting two communicating nodes, such as a \gls{tx} and a \gls{rx}, exhaustive \gls{rt} algorithms typically explore all combinations of object interactions in the scene. For each candidate path, the corresponding ray is traced while ignoring non-interacting objects, and subsequently discarded if it is obstructed or if the computed interaction points fall outside the boundaries of the interacting objects.

According to Fermat's principle, the path taken by a ray between two points makes the traversal time stationary with respect to small variations of the path~\cite{utd-mcnamara}. In homogeneous media, where wave velocity is constant, this reduces to selecting the path whose length is a local extremum. Although this suggests that ray paths can be computed via optimization, the procedure must be computationally efficient, as the number of candidate rays grows exponentially with the number of objects and interactions. Hence, path tracing methods must prioritize speed.

For specular reflection on planar objects, the image method provides exact solutions with computational complexity linear in the number of reflections~\cite{image-method}. However, for diffraction---an essential interaction in wireless communications~\cite{rossiSituMeasurementReflection2000}---the image method does not apply. Instead, minimization-based approaches have been proposed and have demonstrated promising performance~\cite{carluccioEfficientRayTracing2008,puggelliNovelRayTracing2014,vaara-refined}.

Most existing approaches, however, are not designed to scale efficiently on GPUs due to extensive branching, nor are they well-suited for integration with differentiable frameworks. To the best of our knowledge, most existing ray tracers, such as Sionna RT~\cite{hoydis2023sionnartdifferentiableray}, separate the solution into two distinct processes: one for reflection-only paths, and another for reflection-and-diffraction paths. Furthermore, diffraction is usually supported only in a restricted form, often limited to a single diffraction occurring at a fixed position in the interaction chain (typically the last).

In this work, we introduce a generic optimization method capable of handling an arbitrary number of reflections and diffractions in any order, without compromising performance on diffraction-only cases. Moreover, the structure of our problem allows us to compute derivatives efficiently through implicit differentiation, resulting in faster performance than relying on \gls{ad}.

Our main contributions are:
\begin{itemize}
  \item A generic differentiable solver for ray paths with arbitrary reflections and diffractions;
  \item A performance comparison with state-of-the-art methods;
  \item A clean open-source implementation in JAX \cite{jax2018github}, also available through our DiffeRT \gls{rt} tool~\cite{differt}.
\end{itemize}

\section{Related Work}

The fundamental principle underlying ray path determination in free-space is Fermat's principle. For scenarios involving multiple interactions with planar scattering objects, such as planar reflectors and straight edges, the stationary path strictly corresponds to the global minimum of the total length. This naturally leads to a minimization problem whose objective is to find the shortest path between the source and observation points.

Carluccio and Albani~\cite{carluccioEfficientRayTracing2008} presented important work on efficient ray tracing for multiple straight wedge diffraction. Their approach formulates the ray tracing problem as the minimization of the total path length, leveraging the strict convexity of the cost function to guarantee a unique global minimum. The authors used a modified Newton search algorithm with near-quadratic convergence, achieving remarkable computational efficiency for scenarios involving only diffraction interactions. However, their method is specifically designed for edge diffraction and does not address mixed reflection-diffraction scenarios.

Building upon this foundation, Puggelli, Carluccio, and Albani~\cite{puggelliNovelRayTracing2014} extended their approach to handle scenarios comprising both planar reflectors and straight wedges. Their generalized algorithm applies the image method to eliminate reflective surfaces, effectively reducing the problem to a pure diffraction scenario that can be solved using the original Carluccio-Albani method. While this approach maintains computational efficiency, it introduces several limitations for modern parallel computing architectures. Specifically, the treatment of reflections through image theory requires different computational paths depending on the sequence of interaction types, leading to extensive branching operations that are inherently inefficient on GPU architectures, where uniform execution across parallel threads is essential for optimal performance.

Furthermore, the Newton-based approaches in~\cite{carluccioEfficientRayTracing2008,puggelliNovelRayTracing2014} exhibit problem sizes that depend on the number and order of diffractions within the interaction sequence. This variability in problem dimensionality creates additional challenges for GPU implementation, as efficient parallel execution typically requires uniform problem shapes across all threads. The dependency on interaction order also complicates the implementation of vectorized operations, which are crucial for achieving high throughput on modern hardware accelerators.

Despite the computational elegance of Fermat-based path tracing---where paths are determined by minimizing path length according to physical principles---relatively little research has been conducted to improve algorithmic performance or adapt these methods for emerging computational paradigms. Most existing ray tracing frameworks continue to rely on separate algorithms for pure reflection (using the image method) and mixed reflection-diffraction scenarios, leading to code that cannot be efficiently parallelized on GPUs.

Another line of recent research is the advent of \gls{ad} frameworks, which have revolutionized scientific computing by enabling efficient computation of derivatives for complex algorithms~\cite{jaxpaper,drjit}. In the context of \gls{rt}, \gls{ad} capabilities are increasingly important for applications such as inverse design, optimization of wireless networks, and machine learning-based propagation modeling~\cite{hoydis2023sionnartdifferentiableray,inverse,localization}. However, directly applying \gls{ad} to iterative solvers---such as the Newton methods used in existing Fermat-path tracing approaches---often yields suboptimal performance due to the need to differentiate through the entire iteration sequence.

In this work, we propose to bridge these existing contributions by (1) developing a unified framework that handles arbitrary sequences of reflections and diffractions within a single algorithmic procedure that can be accelerated on GPUs, while (2) enabling fast derivative computation through implicit differentiation rather than \gls{ad} through solver iterations. Our approach builds upon the convex optimization foundations established by Carluccio and Albani while addressing the computational architecture requirements of modern ray tracing applications.

\section{Methodology}

The path tracing task consists of finding all feasible paths between two fixed points, namely the \gls{tx} and \gls{rx} antennas, subject to at most $n_\textrm{MAX}$ specular reflection or diffraction interactions with surrounding objects. In practice, we first list all ordered sequences of $n$ objects, corresponding to all candidate paths subject to exactly $n$ interactions, for $n=1, \dots, n_\textrm{MAX}$. For each of these ordered sequences, we find the coordinates of the interaction points that lead to the shortest path (Fermat's principle). The search for the shortest path is formulated as a minimization problem, which we solve using an iterative optimization method. This optimization procedure is detailed hereafter, and is applied in parallel to many different ordered sequences of $n$ objects. As is common in minimization-based approaches~\cite{carluccioEfficientRayTracing2008,puggelliNovelRayTracing2014,mpt-eucap2023}, as well as with the image method~\cite{image-method}, objects are first assumed to be infinitely large when solving for the paths, with possible occlusions by other objects being ignored. The ray paths are then post-processed using standard ray-object intersection tests. This work does not study this post-processing step, as it is common to all methods. Finally, gradients of the solution can be obtained through implicit differentiation.

\subsection{Notation}

We use bold uppercase letters (e.g., $\boldsymbol{A}$) to denote matrices and tensors, bold lowercase letters (e.g., $\boldsymbol{a}$) for vectors, and non-bold lowercase letters (e.g., $a$) for scalars. The $i$-th element (resp. row) of a vector $\boldsymbol{a}$ (resp. matrix $\boldsymbol{A}$) is denoted by $a_i$ (resp. $\boldsymbol{A}_{i}$), while the $(i,j)$-th element of a matrix $\boldsymbol{A}$ is denoted by $a_{i,j}$. The transpose of a vector or matrix is indicated by the superscript $^\top$. The Euclidean norm of a vector $\boldsymbol{a}$ is denoted by $\|\boldsymbol{a}\|$.

\subsection{Problem Formulation}

\begin{figure}
  \centering
  \begin{tikzpicture}[scale=.7,decoration={random steps,segment length=3mm}]
    \pgfmathsetseed{42}
    \node[circle,fill,inner sep=1pt,label=above left:{$\boldsymbol{x}_0$}] (tx) at (0.3,4,-1) {};
    \node[circle,fill,inner sep=1pt,label=below left:{$\boldsymbol{x}_{n+1}$}] (rx) at (12,0,1) {};

    \begin{scope}
      \coordinate (e11) at (0,0,0);
      \coordinate (e12) at (2,0,0);
      \coordinate (e13) at (1,3,0);
      \coordinate (e14) at (2,0,-5);
      \coordinate (e15) at (1,3,-5);
      \coordinate (o1) at (1,3,-1);
      \coordinate (x1) at (1,3,-2);
      \coordinate (e1) at (1,3,-6);

      \filldraw[fill=gray!20] (e11) -- (e13) -- (e15) -- (e14) decorate[rounded corners=1mm] { -- (e12)} decorate[rounded corners=1mm] { -- (e11) };
      \draw (e12) -- (e13);

      \node[circle,fill,inner sep=1pt,label=below right:{$\boldsymbol{b}_1$}] at (o1) {};
      \node[circle,fill,inner sep=1pt,label={[label distance = 2mm]above:{$\boldsymbol{x}_1$}}] at (x1) {};
    \end{scope}

    \begin{scope}[shift={(3,2.2,-3)}]
      \coordinate (p21) at (0,0,0);
      \coordinate (p22) at (0,2,-2);
      \coordinate (p23) at (2,2,-1);
      \coordinate (p24) at (2,0,1);
      \coordinate (o2) at (1,1,-.5);
      \coordinate (x2) at (0.8,0.5,-.5);
      \coordinate (u2) at (1,2.5,-1.5);
      \coordinate (v2) at (-.5,1.25,-.5);
      \coordinate (x3) at (2.5,0.0,+0.8);
      \node[below right] at (x3) {....};

      \filldraw[fill=gray!20,decorate,rounded corners=1mm] (p21) -- (p22) -- (p23) -- (p24) -- cycle;

      \node[circle,fill,inner sep=1pt,label=right:{$\boldsymbol{b}_2$}] at (o2) {};
      \node[circle,fill,inner sep=1pt,label=below:{$\boldsymbol{x}_2$}] at (x2) {};
    \end{scope}

    \begin{scope}[shift={(8,1,+3)}]
      \coordinate (en1) at (0,0,0);
      \coordinate (en2) at (2,0,0);
      \coordinate (en3) at (1,3,0);
      \coordinate (en4) at (2,0,-5);
      \coordinate (en5) at (1,3,-5);
      \coordinate (on) at (1,3,-1);
      \coordinate (xn) at (1,3,-2);
      \coordinate (xnm1) at (0,5,0);
      \coordinate (en) at (1,3,-6);

      \filldraw[fill=gray!20] (en1) -- (en3) -- (en5) -- (en4) decorate[rounded corners=1mm] { -- (en2)} decorate[rounded corners=1mm] { -- (en1) };
      \draw (en2) -- (en3);

      \node[circle,fill,inner sep=1pt,label=below right:{$\boldsymbol{b}_n$}] at (on) {};
      \node[circle,fill,inner sep=1pt,label={[label distance=2mm]above:{$\boldsymbol{x}_n$}}] at (xn) {};
    \end{scope}

    \begin{scope}[dashed,decoration={
          markings,
        mark=at position 1 with {\arrow{latex}}}
      ]
      \draw[postaction={decorate}] (o1) -- (e1) node[above left] {$\boldsymbol{A}_{1,1}$};
      \draw[postaction={decorate}] (o2) -- (u2) node[right] {$\boldsymbol{A}_{2,1}$};
      \draw[postaction={decorate}] (o2) -- (v2) node[above] {$\boldsymbol{A}_{2,2}$};
      \path (x2) -- (x3) node[midway,inner sep=0pt] (midx2x3) {};
      \draw[thick,solid,postaction={decorate}] (x2) -- (midx2x3);
      \draw[thick,dashed] (midx2x3) -- (x3);
      \draw[postaction={decorate}] (on) -- (en) node[above left] {$\boldsymbol{A}_{n,1}$};
    \end{scope}

    \begin{scope}[decoration={
          markings,
        mark=at position 0 with {\arrow{latex}}}
      ]
      \path (xnm1) -- (xn) node[midway,inner sep=0pt] (midxnm1xn) {};
      \draw[thick,dashed] (xnm1) -- (midxnm1xn);
      \draw[thick,solid,postaction={decorate}] (midxnm1xn) -- (xn);
    \end{scope}

    \begin{scope}[thick,decoration={
          markings,
        mark=at position 0.55 with {\arrow{latex}}}
      ]
      \draw[postaction={decorate}] (tx) -- (x1);
      \draw[postaction={decorate}] (x1) -- (x2);
      \draw[postaction={decorate}] (xn) -- (rx);
    \end{scope}

  \end{tikzpicture}
  \caption{Illustration of a ray path with $n$ interactions, including reflections and diffractions, inspired from~\cite[Fig.~1]{carluccioEfficientRayTracing2008}. For conciseness, $\boldsymbol{A}_{i,j}$ is shorthand notation for $\boldsymbol{A}_{i,:,j}$, i.e., the $j$-th base vector of the $i$-th object.}
  \label{fig:setup}
\end{figure}
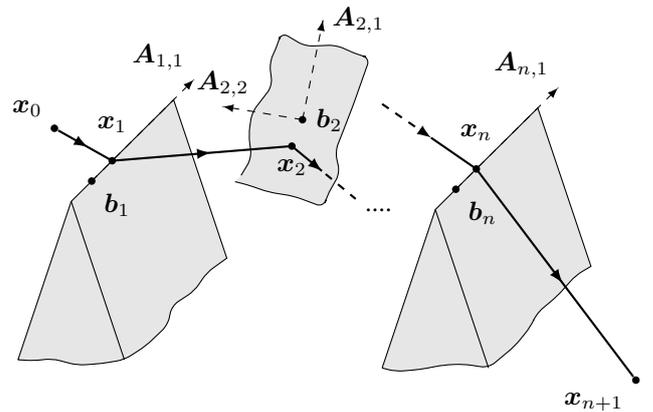

Formally, a ray path $\boldsymbol{X}^*$ is obtained as the solution of the minimization problem
\begin{equation}\label{eq:min}
  \boldsymbol{X}^* = \argmin_{\boldsymbol{X} \in \mathbb{R}^{(n+2)\times 3}} L(\boldsymbol{X}),
\end{equation}
where $L(\boldsymbol{X})$ denotes the total path length
\begin{equation}
  L(\boldsymbol{X}) = \sum_{i=0}^{n} \| \boldsymbol{x}_{i+1} - \boldsymbol{x}_{i} \|,
\end{equation}
with $\boldsymbol{x}_i$, $1 \leq i \leq n$, representing the intermediate interaction points, and $\boldsymbol{x}_{0}$ and $\boldsymbol{x}_{n+1}$ corresponding to the start and end points (typically the \gls{tx} and \gls{rx} coordinates in wireless propagation scenarios), see \autoref{fig:setup}.

In the common case of planar interactions\footnote{Although this may appear restrictive, most 3D models—particularly outdoor urban environments—are described by polygonal objects, which are inherently planar. Detailed non-planar geometries are relatively rare. In~\cite{mpt-eucap2023}, we proposed a minimization-based approach that extends to more complex objects, such as spheres, though at the cost of losing convexity.}, such as specular reflection on planes or diffraction on straight edges, the minimization problem \eqref{eq:min} is strictly convex when expressed in a parametric space~\cite{carluccioEfficientRayTracing2008}.

Indeed, each interaction point $\boldsymbol{x}_i$ can be expressed as an affine transformation of parametric variables $\boldsymbol{t}_i \in \mathbb{R}^{d}$:
\begin{equation}
  \boldsymbol{x}_i = \boldsymbol{A}_i \boldsymbol{t}_i + \boldsymbol{b}_i,
\end{equation}
where $\boldsymbol{b}_i$ is a reference point on the $i$-th planar object, $\boldsymbol{A}_i \in \mathbb{R}^{3\times d}$ is a matrix of basis vectors spanning the object, and $d$ is the intrinsic dimension of the object ($d=2$ for planar reflections, $d=1$ for edge diffractions), see \autoref{fig:setup}.

A key advantage of this formulation is that all matrices $\boldsymbol{A}_i$ (resp. vectors $\boldsymbol{t}_i$ and $\boldsymbol{b}_i$) can be organized into a single tensor $\boldsymbol{A} = \{\boldsymbol{A}_i\}_{i=1}^n$ (resp. matrices $\boldsymbol{T}$ and $\boldsymbol{B}$). When ${d=2}$, interactions corresponding to diffractions are easily handled by using $\boldsymbol{0}$ as the second base vector. While this may increase the apparent number of unknowns for some interaction types, it makes the problem particularly well-suited for massively parallel execution on GPUs, where many ray paths are solved simultaneously. We discuss the effect of this parametrization on convergence speed in later sections.

The minimization problem thus becomes
\begin{equation}\label{eq:min-t}
  \boldsymbol{T}^* = \argmin_{\boldsymbol{T}} L(\boldsymbol{T}; \boldsymbol{A}, \boldsymbol{B}),
\end{equation}
where the objective function is given by:
\begin{equation}
  L(\boldsymbol{T}; \boldsymbol{A}, \boldsymbol{B}) = \sum_{i=0}^{n} \left\| \boldsymbol{A}_{i+1}\boldsymbol{t}_{i+1} + \boldsymbol{b}_{i+1} - \boldsymbol{A}_i\boldsymbol{t}_i - \boldsymbol{b}_i \right\|,
\end{equation}
where $\boldsymbol{A}_0\boldsymbol{t}_0$ and $\boldsymbol{A}_{n+1}\boldsymbol{t}_{n+1}$ are set to zero, and $\boldsymbol{b}_0$ and $\boldsymbol{b}_{n+1}$ are defined as the start and end points, respectively.

\subsection{Iterative Solver}

We utilize the \gls{bfgs} algorithm~\cite[Chapter~3]{bfgs}, a quasi-Newton method, to solve the path length minimization problem \eqref{eq:min-t}. Unlike the custom Newton method used in~\cite{carluccioEfficientRayTracing2008,puggelliNovelRayTracing2014}, \gls{bfgs} does not require computing the Hessian matrix, which can be ill-conditioned in certain scenarios, particularly when interaction points are close together, when the path segments are nearly collinear, or more importantly when one of the base vectors is zero, which occurs for diffraction interactions. Instead, \gls{bfgs} approximates the inverse Hessian $\boldsymbol{H}$ using only gradient evaluations, making it more robust when the Hessian is ill-conditioned.

Unlike traditional \gls{bfgs} implementations, which perform an inexact line search along the descent direction $\boldsymbol{P}$, we use a fixed-point iteration scheme

\begin{equation}\label{eq:fixed-point}
  \alpha^{k+1} = -\frac{\sum_{i=0}^{n}\sfrac{(\Delta \boldsymbol{A}_{i}\boldsymbol{p}_{i})^\top(\Delta\boldsymbol{x}_i)}{\left\|\Delta\boldsymbol{x}_i + \alpha^{k}\Delta \boldsymbol{A}_{i}\boldsymbol{p}_{i} \right\|}}{\sum_{i=0}^{n}\sfrac{(\Delta \boldsymbol{A}_{i}\boldsymbol{p}_{i})^\top(\Delta \boldsymbol{A}_{i}\boldsymbol{p}_{i})}{\left\|\Delta\boldsymbol{x}_i + \alpha^{k}\Delta \boldsymbol{A}_{i}\boldsymbol{p}_{i} \right\|}},
\end{equation}

where $\Delta \boldsymbol{A}_{i}\boldsymbol{p}_{i} = \boldsymbol{A}_{i+1}\boldsymbol{p}_{i+1} - \boldsymbol{A}_{i}\boldsymbol{p}_{i}$, and $\Delta \boldsymbol{x}_{i} = \boldsymbol{x}_{i+1} - \boldsymbol{x}_{i}$, to find the optimal step size $\alpha^*$ that minimizes the objective function along the direction $\boldsymbol{P}$
\begin{equation}
  \alpha^* = \argmin_{\alpha} L(\boldsymbol{T} + \alpha \boldsymbol{P}; \boldsymbol{A}, \boldsymbol{B}).
\end{equation}

In practice, we find that the fixed point iterations converge (in most cases, the right-hand side is locally continuously differentiable and its derivative is smaller than 1 in absolute value). We observe that one iteration is often sufficient to reach the optimal step size with a tolerance of $1\%$.

Finally, and unlike traditional iterative solvers on CPUs, we run the algorithm on a fixed number of iterations, to ensure uniform execution time across all paths in a batch. Indeed, on GPUs, relying on convergence criteria can lead to significant performance degradation as all threads must wait for the slowest one to finish.

\subsection{Implicit Differentiation for Gradient Computation}

\begin{figure*}[t]
  \centering
  \begin{tikzpicture}[every plot/.append style={mark repeat=10},every mark/.append style={opacity=0.7}]
    \pgfplotscreateplotcyclelist{custom black white}{
      blue,every mark/.append style={fill=blue!80!black},mark=*\\
      blue,densely dashed,every mark/.append style={solid,fill=blue!80!black},mark=*\\
      red!20!black,every mark/.append style={fill=red!80!black},mark=otimes*\\
      red!20!black,densely dashed,every mark/.append style={solid,fill=red!80!black},mark=otimes*\\
      brown,every mark/.append style={fill=brown!80!black},mark=square*\\
      gray,every mark/.append style={fill=gray!80!black},mark=diamond*\\
      mark=star\\
    }

    \pgfplotstableread[col sep=comma]{data/perf_diff_1d.txt}\mydataoned;
    \pgfplotstableread[col sep=comma]{data/perf_diff_2d.txt}\mydatatwod;
    \begin{axis}[
        name=plot1a,
        width=0.5\columnwidth,
        height=0.5\columnwidth,
        xtick pos=left,
        xticklabels=\empty,
        ytick={1e-6,1e-3,1e0},
        ymin=1e-7,ymax=1e1,
        xtick={1e0,1e1},
        xmin=4e-1, xmax=3e1,
        xmode=log,
        ymode=log,
        ymajorgrids=true,
        cycle list name=custom black white,
      ]
      \addplot table [x expr=\thisrow{t_ours_1} * 1000,y=e_ours_1] {\mydataoned};
      \addplot table [x expr=\thisrow{t_ours_1} * 1000,y=e_ours_1] {\mydatatwod};
      \addplot table [x expr=\thisrow{t_ours-64_1} * 1000,y=e_ours-64_1] {\mydataoned};
      \addplot table [x expr=\thisrow{t_ours-64_1} * 1000,y=e_ours-64_1] {\mydatatwod};
      \addplot table [x expr=\thisrow{t_gd_1} * 1000,y=e_gd_1] {\mydataoned};
      \addplot table [x expr=\thisrow{t_malbani_1} * 1000,y=e_malbani_1] {\mydataoned};
      \addplot table [x expr=\thisrow{t_l-bfgs_1} * 1000,y=e_l-bfgs_1] {\mydataoned};
    \end{axis}
    \begin{axis}[
        name=plot2a,
        at={($ (.1cm,0cm) + (plot1a.south east) $)},
        width=0.5\columnwidth,
        height=0.5\columnwidth,
        xtick pos=left,
        xticklabels=\empty,
        yticklabels=\empty,
        ytick={1e-6,1e-3,1e0},
        ymin=1e-7,ymax=1e1,
        xtick={1e0,1e1},
        xmin=4e-1, xmax=3e1,
        xmode=log,
        ymode=log,
        ymajorgrids=true,
        cycle list name=custom black white,
      ]
      \addplot table [x expr=\thisrow{t_ours_2} * 1000,y=e_ours_2] {\mydataoned};
      \addplot table [x expr=\thisrow{t_ours_2} * 1000,y=e_ours_2] {\mydatatwod};
      \addplot table [x expr=\thisrow{t_ours-64_2} * 1000,y=e_ours-64_2] {\mydataoned};
      \addplot table [x expr=\thisrow{t_ours-64_2} * 1000,y=e_ours-64_2] {\mydatatwod};
      \addplot table [x expr=\thisrow{t_gd_2} * 1000,y=e_gd_2] {\mydataoned};
      \addplot table [x expr=\thisrow{t_malbani_2} * 1000,y=e_malbani_2] {\mydataoned};
      \addplot table [x expr=\thisrow{t_l-bfgs_2} * 1000,y=e_l-bfgs_2] {\mydataoned};
    \end{axis}
    \begin{axis}[
        name=plot3a,
        at={($ (.1cm,0cm) + (plot2a.south east) $)},
        width=0.5\columnwidth,
        height=0.5\columnwidth,
        xtick pos=left,
        xticklabels=\empty,
        yticklabels=\empty,
        ytick={1e-6,1e-3,1e0},
        ymin=1e-7,ymax=1e1,
        xtick={1e0,1e1},
        xmin=4e-1, xmax=3e1,
        xmode=log,
        ymode=log,
        legend style={font=\small, at={(0.5,1.025)},anchor=south,name=leg,nodes={anchor=base}},
        ymajorgrids=true,
        cycle list name=custom black white,
        legend columns=-1,legend style={column sep=1ex},
        legend image post style={yshift=.5ex},
        every legend to name picture/.style={
          baseline={(leg.base)},
        },
      ]
      \addlegendimage{blue,only marks,fill=blue!80!black,mark=*}
      \addlegendimage{red!20!black,only marks,fill=red!80!black,mark=otimes*}
      \addlegendimage{no marks,draw=black!80!white,opacity=0.5,ultra thick,dashed}
      \addlegendimage{brown,only marks,fill=brown!80,mark=square*}
      \addlegendimage{gray,only marks,fill=gray!80!black,mark=diamond*}
      \addlegendimage{only marks,fill=black!20,mark=star}
      \legend{ours,ours-64,IM,GD,CA,L-BFGS}
      \addplot table [x expr=\thisrow{t_ours_3} * 1000,y=e_ours_3] {\mydataoned};
      \addplot table [x expr=\thisrow{t_ours_3} * 1000,y=e_ours_3] {\mydatatwod};
      \addplot table [x expr=\thisrow{t_ours-64_3} * 1000,y=e_ours-64_3] {\mydataoned};
      \addplot table [x expr=\thisrow{t_ours-64_3} * 1000,y=e_ours-64_3] {\mydatatwod};
      \addplot table [x expr=\thisrow{t_gd_3} * 1000,y=e_gd_3] {\mydataoned};
      \addplot table [x expr=\thisrow{t_malbani_3} * 1000,y=e_malbani_3] {\mydataoned};
      \addplot table [x expr=\thisrow{t_l-bfgs_3} * 1000,y=e_l-bfgs_3] {\mydataoned};
    \end{axis}
    \begin{axis}[
        name=plot4a,
        at={($ (.1cm,0cm) + (plot3a.south east) $)},
        width=0.5\columnwidth,
        height=0.5\columnwidth,
        xtick pos=left,
        xticklabels=\empty,
        yticklabels=\empty,
        ytick={1e-6,1e-3,1e0},
        ymin=1e-7,ymax=1e1,
        xtick={1e0,1e1},
        xmin=4e-1, xmax=3e1,
        xmode=log,
        ymode=log,
        ymajorgrids=true,
        cycle list name=custom black white,
      ]
      \addplot table [x expr=\thisrow{t_ours_4} * 1000,y=e_ours_4] {\mydataoned};
      \addplot table [x expr=\thisrow{t_ours_4} * 1000,y=e_ours_4] {\mydatatwod};
      \addplot table [x expr=\thisrow{t_ours-64_4} * 1000,y=e_ours-64_4] {\mydataoned};
      \addplot table [x expr=\thisrow{t_ours-64_4} * 1000,y=e_ours-64_4] {\mydatatwod};
      \addplot table [x expr=\thisrow{t_gd_4} * 1000,y=e_gd_4] {\mydataoned};
      \addplot table [x expr=\thisrow{t_malbani_4} * 1000,y=e_malbani_4] {\mydataoned};
      \addplot table [x expr=\thisrow{t_l-bfgs_4} * 1000,y=e_l-bfgs_4] {\mydataoned};
    \end{axis}
    \begin{axis}[
        name=plot5a,
        at={($ (.1cm,0cm) + (plot4a.south east) $)},
        width=0.5\columnwidth,
        height=0.5\columnwidth,
        xtick pos=left,
        xticklabels=\empty,
        yticklabels=\empty,
        ytick={1e-6,1e-3,1e0},
        ymin=1e-7,ymax=1e1,
        xtick={1e0,1e1},
        xmin=4e-1, xmax=3e1,
        xmode=log,
        ymode=log,
        ymajorgrids=true,
        cycle list name=custom black white,
      ]
      \addplot table [x expr=\thisrow{t_ours_5} * 1000,y=e_ours_5] {\mydataoned};
      \addplot table [x expr=\thisrow{t_ours_5} * 1000,y=e_ours_5] {\mydatatwod};
      \addplot table [x expr=\thisrow{t_ours-64_5} * 1000,y=e_ours-64_5] {\mydataoned};
      \addplot table [x expr=\thisrow{t_ours-64_5} * 1000,y=e_ours-64_5] {\mydatatwod};
      \addplot table [x expr=\thisrow{t_gd_5} * 1000,y=e_gd_5] {\mydataoned};
      \addplot table [x expr=\thisrow{t_malbani_5} * 1000,y=e_malbani_5] {\mydataoned};
      \addplot table [x expr=\thisrow{t_l-bfgs_5} * 1000,y=e_l-bfgs_5] {\mydataoned};
    \end{axis}
    \node[draw,fill=black!10,anchor=north east,font=\small] at ($ (-.1cm,-.1cm) + (plot1a.north east) $) {$n=1$};
    \node[draw,fill=black!10,anchor=north east,font=\small] at ($ (-.1cm,-.1cm) + (plot2a.north east) $) {$n=2$};
    \node[draw,fill=black!10,anchor=north east,font=\small] at ($ (-.1cm,-.1cm) + (plot3a.north east) $) {$n=3$};
    \node[draw,fill=black!10,anchor=north east,font=\small] at ($ (-.1cm,-.1cm) + (plot4a.north east) $) {$n=4$};
    \node[draw,fill=black!10,anchor=north east,font=\small] at ($ (-.1cm,-.1cm) + (plot5a.north east) $) {$n=5$};

    \pgfplotstableread[col sep=comma]{data/perf_refl_2d.txt}\mydataoned;
    \pgfplotstableread[col sep=comma]{data/perf_mixed_2d.txt}\mydatatwod;
    \begin{axis}[
        name=plot1b,
        anchor=north,
        at={($ (0cm,-.1cm) + (plot1a.south) $)},
        width=0.5\columnwidth,
        height=0.5\columnwidth,
        xtick pos=left,
        ytick={1e-4,1e-2,1e0},
        ymin=1e-5,ymax=1e1,
        xtick={1e0,1e1},
        xticklabels={1,10},
        xmin=4e-1, xmax=3e1,
        xmode=log,
        ymode=log,
        ymajorgrids=true,
        cycle list name=custom black white,
      ]
      \addplot table [x expr=\thisrow{t_ours_1} * 1000,y=e_ours_1] {\mydataoned};
      \addplot table [x expr=\thisrow{t_ours_1} * 1000,y=e_ours_1] {\mydatatwod};
      \addplot table [x expr=\thisrow{t_ours-64_1} * 1000,y=e_ours-64_1] {\mydataoned};
      \addplot table [x expr=\thisrow{t_ours-64_1} * 1000,y=e_ours-64_1] {\mydatatwod};
      \addplot table [x expr=\thisrow{t_gd_1} * 1000,y=e_gd_1] {\mydataoned};
      \addplot table [x expr=\thisrow{t_malbani_1} * 1000,y=e_malbani_1] {\mydataoned};
      \addplot table [x expr=\thisrow{t_l-bfgs_1} * 1000,y=e_l-bfgs_1] {\mydataoned};
      \draw[black!80!white,opacity=0.5,ultra thick,dashed] (current axis.south-|0.454, 0) -- (current axis.north-|0.454,0);
    \end{axis}
    \begin{axis}[
        name=plot2b,
        at={($ (.1cm,0cm) + (plot1b.south east) $)},
        width=0.5\columnwidth,
        height=0.5\columnwidth,
        xtick pos=left,
        yticklabels=\empty,
        ytick={1e-4,1e-2,1e0},
        ymin=1e-5,ymax=1e1,
        xtick={1e0,1e1},
        xticklabels={1,10},
        xmin=4e-1, xmax=3e1,
        xmode=log,
        ymode=log,
        ymajorgrids=true,
        cycle list name=custom black white,
      ]
      \addplot table [x expr=\thisrow{t_ours_2} * 1000,y=e_ours_2] {\mydataoned};
      \addplot table [x expr=\thisrow{t_ours_2} * 1000,y=e_ours_2] {\mydatatwod};
      \addplot table [x expr=\thisrow{t_ours-64_2} * 1000,y=e_ours-64_2] {\mydataoned};
      \addplot table [x expr=\thisrow{t_ours-64_2} * 1000,y=e_ours-64_2] {\mydatatwod};
      \addplot table [x expr=\thisrow{t_gd_2} * 1000,y=e_gd_2] {\mydataoned};
      \addplot table [x expr=\thisrow{t_malbani_2} * 1000,y=e_malbani_2] {\mydataoned};
      \addplot table [x expr=\thisrow{t_l-bfgs_2} * 1000,y=e_l-bfgs_2] {\mydataoned};
      \draw[black!80!white,opacity=0.5,ultra thick,dashed] (current axis.south-|0.492, 0) -- (current axis.north-|0.492,0);
    \end{axis}
    \begin{axis}[
        name=plot3b,
        at={($ (.1cm,0cm) + (plot2b.south east) $)},
        width=0.5\columnwidth,
        height=0.5\columnwidth,
        xlabel={Execution time (ms)},
        xtick pos=left,
        yticklabels=\empty,
        ytick={1e-4,1e-2,1e0},
        ymin=1e-5,ymax=1e1,
        xtick={1e0,1e1},
        xticklabels={1,10},
        xmin=4e-1, xmax=3e1,
        xmode=log,
        ymode=log,
        ymajorgrids=true,
        cycle list name=custom black white,
      ]
      \addplot table [x expr=\thisrow{t_ours_3} * 1000,y=e_ours_3] {\mydataoned};
      \addplot table [x expr=\thisrow{t_ours_3} * 1000,y=e_ours_3] {\mydatatwod};
      \addplot table [x expr=\thisrow{t_ours-64_3} * 1000,y=e_ours-64_3] {\mydataoned};
      \addplot table [x expr=\thisrow{t_ours-64_3} * 1000,y=e_ours-64_3] {\mydatatwod};
      \addplot table [x expr=\thisrow{t_gd_3} * 1000,y=e_gd_3] {\mydataoned};
      \addplot table [x expr=\thisrow{t_malbani_3} * 1000,y=e_malbani_3] {\mydataoned};
      \addplot table [x expr=\thisrow{t_l-bfgs_3} * 1000,y=e_l-bfgs_3] {\mydataoned};
      \draw[black!80!white,opacity=0.5,ultra thick,dashed] (current axis.south-|0.508, 0) -- (current axis.north-|0.508,0);
    \end{axis}
    \begin{axis}[
        name=plot4b,
        at={($ (.1cm,0cm) + (plot3b.south east) $)},
        width=0.5\columnwidth,
        height=0.5\columnwidth,
        xtick pos=left,
        yticklabels=\empty,
        ytick={1e-4,1e-2,1e0},
        ymin=1e-5,ymax=1e1,
        xtick={1e0,1e1},
        xticklabels={1,10},
        xmin=4e-1, xmax=3e1,
        xmode=log,
        ymode=log,
        ymajorgrids=true,
        cycle list name=custom black white,
      ]
      \addplot table [x expr=\thisrow{t_ours_4} * 1000,y=e_ours_4] {\mydataoned};
      \addplot table [x expr=\thisrow{t_ours_4} * 1000,y=e_ours_4] {\mydatatwod};
      \addplot table [x expr=\thisrow{t_ours-64_4} * 1000,y=e_ours-64_4] {\mydataoned};
      \addplot table [x expr=\thisrow{t_ours-64_4} * 1000,y=e_ours-64_4] {\mydatatwod};
      \addplot table [x expr=\thisrow{t_gd_4} * 1000,y=e_gd_4] {\mydataoned};
      \addplot table [x expr=\thisrow{t_malbani_4} * 1000,y=e_malbani_4] {\mydataoned};
      \addplot table [x expr=\thisrow{t_l-bfgs_4} * 1000,y=e_l-bfgs_4] {\mydataoned};
      \draw[black!80!white,opacity=0.5,ultra thick,dashed] (current axis.south-|0.524, 0) -- (current axis.north-|0.524,0);
    \end{axis}
    \begin{axis}[
        name=plot5b,
        at={($ (.1cm,0cm) + (plot4b.south east) $)},
        width=0.5\columnwidth,
        height=0.5\columnwidth,
        xtick pos=left,
        yticklabels=\empty,
        ytick={1e-4,1e-2,1e0},
        ymin=1e-5,ymax=1e1,
        xtick={1e0,1e1},
        xticklabels={1,10},
        xmin=4e-1, xmax=3e1,
        xmode=log,
        ymode=log,
        ymajorgrids=true,
        cycle list name=custom black white,
      ]
      \addplot table [x expr=\thisrow{t_ours_5} * 1000,y=e_ours_5] {\mydataoned};
      \addplot table [x expr=\thisrow{t_ours_5} * 1000,y=e_ours_5] {\mydatatwod};
      \addplot table [x expr=\thisrow{t_ours-64_5} * 1000,y=e_ours-64_5] {\mydataoned};
      \addplot table [x expr=\thisrow{t_ours-64_5} * 1000,y=e_ours-64_5] {\mydatatwod};
      \addplot table [x expr=\thisrow{t_gd_5} * 1000,y=e_gd_5] {\mydataoned};
      \addplot table [x expr=\thisrow{t_malbani_5} * 1000,y=e_malbani_5] {\mydataoned};
      \addplot table [x expr=\thisrow{t_l-bfgs_5} * 1000,y=e_l-bfgs_5] {\mydataoned};
      \draw[black!80!white,opacity=0.5,ultra thick,dashed] (current axis.south-|0.544, 0) -- (current axis.north-|0.544,0);
    \end{axis}
    \node[draw,fill=black!10,anchor=south west,font=\small] at ($ (+.1cm,+.1cm) + (plot1b.south west) $) {$n=1$};
    \node[draw,fill=black!10,anchor=south west,font=\small] at ($ (+.1cm,+.1cm) + (plot2b.south west) $) {$n=2$};
    \node[draw,fill=black!10,anchor=south west,font=\small] at ($ (+.1cm,+.1cm) + (plot3b.south west) $) {$n=3$};
    \node[draw,fill=black!10,anchor=south west,font=\small] at ($ (+.1cm,+.1cm) + (plot4b.south west) $) {$n=4$};
    \node[draw,fill=black!10,anchor=south west,font=\small] at ($ (+.1cm,+.1cm) + (plot5b.south west) $) {$n=5$};

    \path (plot1a.north west) -- (plot1b.south) coordinate[pos=0.5] (ylabel);
    \path (ylabel) -- ++(-45pt,0cm) node[rotate=90,anchor=south] {Average error on $\boldsymbol{X}^*$};
  \end{tikzpicture}
  \caption{Average error on 1000 paths against time taken by different solvers, for $n$ increasing from 1 to 5. Top row corresponds to 1D problems (diffractions), bottom row to 2D problems (reflections). Diffractions solved with $d=2$ are shown with dashed lines in the top row, only with our solver as other methods fail. Similarly, mixed reflections and diffractions cases are show in dashed in the bottom row, for our solver only. Our solver is run in two configurations: standard (ours) and with 64 fixed-point iterations per step (ours-64). Because the image method is exact, a vertical line is shown to indicate its execution time.}
  \label{fig:perf}
\end{figure*}
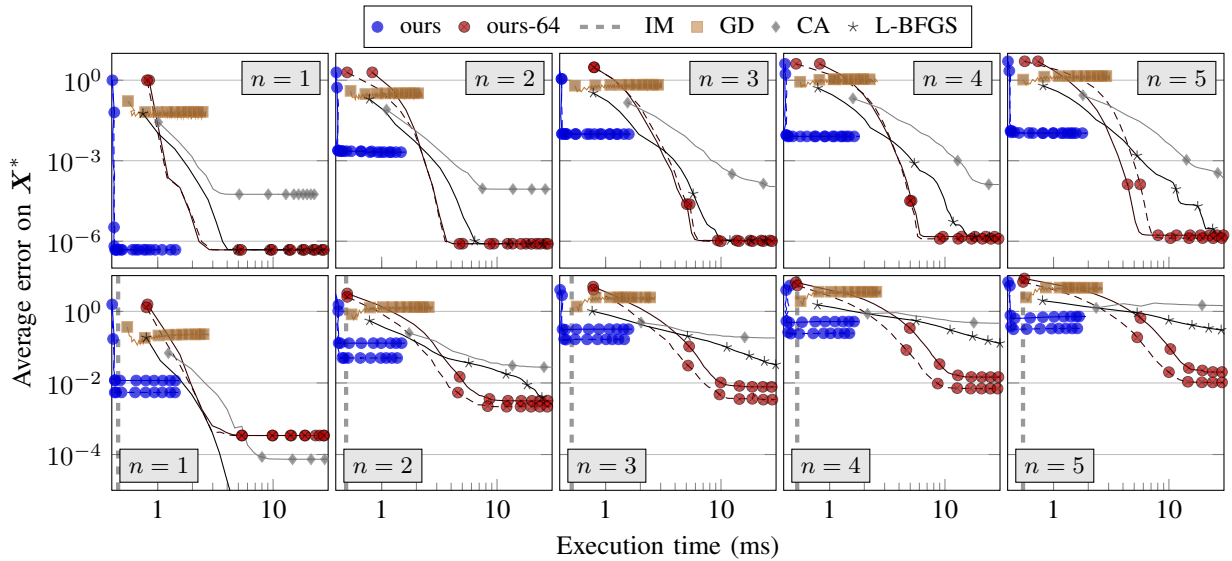

In many applications, such as inverse problems or parameter optimization, it is necessary to compute the gradient of one scalar quantity (e.g., the received power)  with respect to many problem parameters (e.g., object positions or orientations). A straightforward approach is to use reverse-mode \gls{ad} to differentiate through all the iterations of the solver. However, this can be computationally expensive and memory-intensive, especially when the direct method (here, radio propagation via \gls{rt}) involves many iterations, as \gls{ad} must record and backpropagate through every operation in the iterative process.

To address this, we leverage the fact that, at convergence, our solver produces a solution $\boldsymbol{T}^*(\boldsymbol{\theta})$ (with $\boldsymbol{\theta} = (\boldsymbol{A}, \boldsymbol{B})$) that satisfies the first-order optimality condition:

\begin{equation}\label{eq:optimality}
  \nabla_{\boldsymbol{T}} L(\boldsymbol{T}^*(\boldsymbol{\theta}); \boldsymbol{\theta}) = \boldsymbol{0},
\end{equation}
where $L$ is the objective function.

Assuming the solver has converged to such a stationary point, we can use the implicit function theorem~\cite[Theorem~7-6, p.~146]{apostol} to compute the total derivative of $\boldsymbol{T}^*$ with respect to the parameters $\boldsymbol{\theta}$, without differentiating through the entire sequence of solver steps. However, we still rely on \gls{ad} to compute the Jacobian vectors of $\nabla_{\boldsymbol{T}} L$.

Applying the implicit function theorem to \eqref{eq:optimality} yields:
\begin{equation} \frac{\partial}{\partial \boldsymbol{\theta}} \nabla_{\boldsymbol{T}} L(\boldsymbol{T}^*(\boldsymbol{\theta}); \boldsymbol{\theta}) + \frac{\partial \nabla_{\boldsymbol{T}} L}{\partial \boldsymbol{T}} \frac{\partial \boldsymbol{T}^*}{\partial \boldsymbol{\theta}} = \boldsymbol{0}.
\end{equation}

Rearranging, we obtain:
\begin{equation} \frac{\partial \boldsymbol{T}^*}{\partial \boldsymbol{\theta}} = -\left[\frac{\partial \nabla_{\boldsymbol{T}} L}{\partial \boldsymbol{T}}\right]^{-1} \frac{\partial}{\partial \boldsymbol{\theta}} \nabla_{\boldsymbol{T}} L.
\end{equation}

In practice, we are often interested in computing vector-Jacobian products of the form $\boldsymbol{v}^\top \frac{\partial \boldsymbol{T}^*}{\partial \boldsymbol{\theta}}$, as required by reverse-mode AD. This can be done efficiently by solving the following linear system:

\begin{equation} \boldsymbol{u}^\top = -\boldsymbol{v}^\top \left[\frac{\partial \nabla_{\boldsymbol{T}} L}{\partial \boldsymbol{T}}\right]^{-1},
\end{equation}

or equivalently

\begin{equation} \left[\frac{\partial \nabla_{\boldsymbol{T}} L}{\partial \boldsymbol{T}}\right]^\top\boldsymbol{u} = -\boldsymbol{v},
\end{equation}

where $\frac{\partial \nabla_{\boldsymbol{T}} L}{\partial \boldsymbol{T}}$ is a symmetric positive semi-definite matrix, being the Hessian of the convex function $L$, and then computing

\begin{equation} \boldsymbol{v}^\top \frac{\partial \boldsymbol{T}^*}{\partial \boldsymbol{\theta}} = \boldsymbol{u}^\top \frac{\partial}{\partial \boldsymbol{\theta}} \nabla_{\boldsymbol{T}} L.
\end{equation}

This approach avoids differentiating through the entire solver, significantly reducing computational cost and memory usage, while providing exact gradients under the assumption of solver convergence.

\section{Simulation Results}

We evaluate our method on a variety of random scenarios to demonstrate its effectiveness and computational efficiency. All simulations are conducted on an NVIDIA GeForce RTX 3070 GPU with 8 GB memory. We implemented the algorithm with JAX~\cite{jax2018github}, leveraging its just-in-time compilation and convenient support for generating GPU-compatible code. While JAX provides built-in \gls{ad} capabilities, it also allows defining custom derivative rules, so we can implement implicit differentiation manually to avoid differentiating through solver iterations in backward-mode \gls{ad}.%

\subsection{Performance Comparison}

We benchmark our approach against the following solvers:
\begin{itemize}
  \item the image method (IM), in reflection-only scenarios;
  \item a standard gradient descent (GD);
  \item the Carluccio and Albani (CA) method~\cite{carluccioEfficientRayTracing2008}, extended to $d=2$ when relevant;
  \item and a limited-memory \gls{bfgs} (L-BFGS) solver implemented in the Optax library~\cite{optax}.
\end{itemize}

Each solver processes $1000$ ray paths in parallel, performing up to $100$ iterations. We record the average error and execution time using single-precision floating-point (32-bit) arithmetic. The average error is defined as the mean Euclidean distance between the estimated interaction points and the ground-truth points obtained from a high-precision solver on a CPU.

\autoref{fig:perf} summarizes the results for increasing path complexity, with the number of interactions $n$ ranging from 1 to 5. The top row corresponds to diffraction-only problems ($d=1$), and the bottom row to reflection-only problems ($d=2$). Dashed lines indicate cases involving diffractions represented using $d=2$ instead of $d=1$, or mixed reflections and diffractions ($d=2$), which are only supported by our method.

As shown in \autoref{fig:perf}, the standard gradient descent (GD) consistently fails to converge. In diffraction-only configurations ($d=1$), our solver achieves both the lowest error and the fastest convergence across all cases. For $n>1$, increasing the number of fixed-point iterations enables convergence down to machine precision. Notably, extending the optimization to $d=2$ has negligible impact on accuracy or speed for our method, as the $d=1$ and $d=2$ curves almost coincide, while other solvers fail to handle the additional dimension.%

In reflection-only scenarios, a similar trend is observed. Except for the simplest case ($n=1$), where L-BFGS reaches machine precision, our solver consistently delivers higher accuracy in less time. Increasing the number of fixed-point iterations again improves precision for larger $n$. The image method remains the most efficient, achieving comparable or better accuracy at an order-of-magnitude lower runtime.

Finally, in mixed reflection-diffraction configurations (dashed curves in the bottom row of \autoref{fig:perf}), the average convergence is even faster than in reflection-only cases, likely because diffractions introduce only one unknown per interaction, whereas reflections introduce two.

\subsection{Implicit vs Automatic Differentiation}

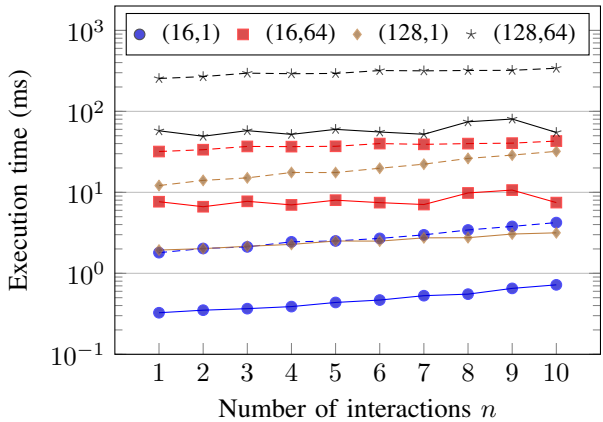
\begin{figure}
  \centering
  \begin{tikzpicture}[every mark/.append style={opacity=0.7}]
    \pgfplotscreateplotcyclelist{custom black white}{
      blue,every mark/.append style={fill=blue!80!black},mark=*\\
      blue,densely dashed,every mark/.append style={solid,fill=blue!80!black},mark=*\\
      red,every mark/.append style={fill=red!80!black},mark=square*\\
      red,densely dashed,every mark/.append style={solid,fill=red!80!black},mark=square*\\
      brown,every mark/.append style={fill=brown!80!black},mark=diamond*\\
      brown,densely dashed,every mark/.append style={solid,fill=brown!80!black},mark=diamond*\\
      mark=star\\
      densely dashed,every mark/.append style={solid},mark=star\\
    }
    \begin{axis}[
        width=0.9\columnwidth,
        height=0.7\columnwidth,
        xlabel={Number of interactions $n$},
        ylabel={Execution time (ms)},
        xtick pos=left,
        xmin=0, xmax=11,
        ymin=1e-1,ymax=2e3,
        xtick={1,2,3,4,5,6,7,8,9,10},
        ymode=log,
        legend style={font=\small, at={(0.5,0.98)},anchor=north,name=leg,nodes={anchor=base}},
        ymajorgrids=true,
        cycle list name=custom black white,
        legend columns=-1,legend style={column sep=1ex},
        legend image post style={yshift=.5ex},
        every legend to name picture/.style={
          baseline={(leg.base)},
        },
      ]
      \pgfplotstableread[col sep=comma]{data/auto_vs_impl_diff.txt}\mydata;
      \addplot table [x=num_inter,y expr=\thisrow{impl-16-1-2} * 1000] {\mydata};
      \addplot table [x=num_inter,y expr=\thisrow{ad-16-1-2} * 1000] {\mydata};
      \addplot table [x=num_inter,y expr=\thisrow{impl-16-64-2} * 1000] {\mydata};
      \addplot table [x=num_inter,y expr=\thisrow{ad-16-64-2} * 1000] {\mydata};
      \addplot table [x=num_inter,y expr=\thisrow{impl-128-1-2} * 1000] {\mydata};
      \addplot table [x=num_inter,y expr=\thisrow{ad-128-1-2} * 1000] {\mydata};
      \addplot table [x=num_inter,y expr=\thisrow{impl-128-64-2} * 1000] {\mydata};
      \addplot table [x=num_inter,y expr=\thisrow{ad-128-64-2} * 1000] {\mydata};
      \addlegendimage{black,only marks,fill=blue!80!black,mark=*}
      \addlegendimage{red,only marks,fill=red!80!black,mark=square*}
      \addlegendimage{brown,only marks,fill=brown!80!black,mark=diamond*}
      \addlegendimage{only marks,mark=star}
      \legend{,,,,,,,,(16,1),(16,64),(128,1),(128,64)}
    \end{axis}
  \end{tikzpicture}
  \caption{Execution time for computing the gradient of 1000 path lengths at $\boldsymbol{T}^*(\boldsymbol{\theta})$ with respect to all object parameters, i.e., $\nabla_{\boldsymbol{\theta}} \boldsymbol{L}(\boldsymbol{T}^*(\boldsymbol{\theta});\boldsymbol{\theta})$, using implicit differentiation (solid lines) and automatic differentiation (dashed lines), as a function of the number of interactions $n$, for $d=2$. The solver is run for 16 or 128 iterations, with 1 or 64 fixed-point iterations per step.}
  \label{fig:impl-vs-auto}
\end{figure}

When comparing the execution time of gradient computation using implicit differentiation versus \gls{ad}, we observe significant performance improvements with our approach (nearly 10 times faster) that remain constant regardless of the number of solver iterations, the number of fixed-point iterations per step, or the number of considered interactions; see \autoref{fig:impl-vs-auto}. We also observed very similar results for $d=1$.%

\section{Conclusion and Future Work}

We have presented a unified, differentiable ray tracing method that reformulates path finding as a convex optimization problem and leverages implicit differentiation for efficient gradient computation. The proposed approach handles arbitrary sequences of reflections and diffractions within a single optimization framework, eliminating branching complexity and enabling efficient parallel execution on GPUs.

Simulation results demonstrate that the solver achieves fast and accurate convergence across a wide range of interaction scenarios, outperforming existing minimization-based approaches in both speed and robustness. Nevertheless, the results also reveal certain limitations, including slower convergence than the image method in reflection-only cases.

Future work will focus on reformulating the current nonsmooth optimization problem into a second order cone program, to be solved with dedicated solvers~\cite{ecos,scs} to improve convergence properties while preserving the advantages of implicit differentiation~\cite{impl}. Additional efforts will aim at refining the selection of initial candidates, improving line-search strategies, and analyzing the convergence behavior of the proposed fixed-point scheme.

\balance

\section*{Acknowledgment}

Part of this research was funded by the FRS-FNRS (PDR, QuadSense, T.0160.24).

\bibliographystyle{IEEEtran}
\bibliography{IEEEabrv,biblio}

\end{document}